\RequirePackage{fix-cm}
\documentclass[smallextended]{svjour3}       
\smartqed  
\usepackage{appendix}
\usepackage{amsmath}
\usepackage{graphicx}
\usepackage{lineno}
\usepackage{array}
\usepackage{longtable}
\usepackage{mathptmx}
\usepackage{physics}
\usepackage{amsmath}
\usepackage{amssymb}
\usepackage{bbm}
\usepackage{color}
\usepackage{mathtools}
\usepackage{makecell}
\usepackage{bm}
\usepackage{todonotes}
\usepackage{diagbox}
\usepackage{tablefootnote}
\usepackage[ruled,vlined,linesnumbered]{algorithm2e}
\usepackage{graphicx}
 \usepackage{booktabs}
\usepackage{verbatim}
\usepackage[english]{babel}
\usepackage{csquotes} 
\usepackage[margin=1in]{geometry}
\usepackage{xfrac}
\usepackage{enumitem}

\usepackage[hidelinks]{hyperref}
\usepackage{thm-restate}

\newcommand*\patchAmsMathEnvironmentForLineno[1]{%
\expandafter\let\csname old#1\expandafter\endcsname\csname #1\endcsname
\expandafter\let\csname oldend#1\expandafter\endcsname\csname end#1\endcsname
\renewenvironment{#1}%
{\linenomath\csname old#1\endcsname}%
{\csname oldend#1\endcsname\endlinenomath}}%
\newcommand*\patchBothAmsMathEnvironmentsForLineno[1]{%
\patchAmsMathEnvironmentForLineno{#1}%
\patchAmsMathEnvironmentForLineno{#1*}}%
\AtBeginDocument{%
\patchBothAmsMathEnvironmentsForLineno{equation}%
\patchBothAmsMathEnvironmentsForLineno{align}%
\patchBothAmsMathEnvironmentsForLineno{flalign}%
\patchBothAmsMathEnvironmentsForLineno{alignat}%
\patchBothAmsMathEnvironmentsForLineno{gather}%
\patchBothAmsMathEnvironmentsForLineno{multline}%
}

\begin{document}

\title{Quantum Multiplayer Colonel Blotto Game
}


\author{ Joydeep Naskar$^{1}$         \and
        Alan C. Maioli  $^{2*}$ 
}


\institute{      $^1$School of Physical Sciences, National Institute of Science Education and Research. \at P.O. Jatni, Khurda, Odisha, India. PIN- 752050 \\
              \email{naskarjoydeep@gmail.com}  
              \\ $^2$ \email{alanmaioli90@gmail.com}
\newline *Corresponding Author 
}

\date{Received: 01 August 2020}

\maketitle

\begin{abstract}
In this work we successfully present a quantum version of the multiplayer Colonel Blotto game. We find that players with access to the quantum strategies has a advantage over the classical ones. The payoff is invariant under the order of the operator's strategies.

\end{abstract}

\section{Introduction}
\label{intro}

Quantum game theory is a branch of quantum information which overlaps with game theory. The last one is an active field of mathematics with a plethora of works and applications. Nowadays, physicists refer games from game theory as \textit{classical games}. One can rewrite a classical game using state vectors and operators to depict strategies and measurement to obtain the payoff. Along these lines, one can add new operators without classic analogue, and with entanglement and superposition \cite{Meyer} new features appear. This process is often called \textit{quantization of a game}. Therefore, any quantum game must preserve all features from its classical counterpart. Since 1999 several classical games was brought to the quantum realm due to the intrinsic probabilistic nature of quantum theory and game theory (for mixed strategies), such as Prisoners' Dilemma \cite{Eisert}, Hawk-Dove \cite{Ahmad,Tomassini}, Quantum duel \cite{Flitney_2004,Schmidt_2012} and recently Colonel Blotto \cite{maioli2019quantization}. The quantum games are investigated in order to find new properties \cite{santos2020entanglement,dhiman2020implementation,anand2020solving,faleiro2020quantum,du2003multiprisoner} or deep connections with game theory \cite{brunner2013connection}.\\
Our purpose is to present a quantum version of multiplayer Colonel Blotto game. Therefore we use a similar procedure made by the Maioli \textit{et al} \cite{maioli2019quantization}, however there is a subtle difference near the end of the game. In other words, we use the density matrix formalism in order to perform the measurements and obtain the payoff. This choice has the advantage to separate the subsystems via partial trace. One can find similar use of the density matrix formalism in quantum games in the reference \cite{guo2008survey}, it also contains an excellent description of quantum games. In the construction by Maioli et al in \cite{maioli2019quantization}, 2 players, namely Blotto and enemy shared a common qubit for representation of their soldiers. This led to a philosophical discussion introduced by \cite{naskarcomment2020}. In the following construction, we have resolved that conundrum.\\
A classical description of the game is given at section (\ref{sec: classicalgame}). In this work we successfully present a quantum version of the multiplayer Colonel Blotto game on section (\ref{sec: quantumgame}), and this is corroborated by the examples on subsection (\ref{subsec: classicalstrategies}). Therefore, we investigate the payoff's alteration due to the existence of quantum strategies at subsection (\ref{subsec: quantummove}).

\section{The Classical Game}\label{sec: classicalgame}
The Colonel Blotto game can be formulated in different ways\cite{genblotto2020,discblotto2008,contblotto1950,jayantimultiplayerblotto2020}. In this section we present a classical version of the multiplayer Colonel Blotto game, a similar version of the game can be found at \cite{jayantimultiplayerblotto2020}. We consider the number of players $N>2$ and it has $n$ number of battlefields. Player 1 is defined as Blotto, while other players are the enemies, player 2 is enemy 1 and so on. Any strategy chosen by player $j$ is presented as a vector,
\begin{equation}\label{eq: estrategyclassicgame}
    \mathbf{x}_j= (x_1^j,x_2^j,...,x_n^j),
\end{equation}
where $x_k^j$ is the number of soldiers the player $j$ allocate on the battlefield $k$. We are considering a continuous game \cite{contblotto1950}, so the numbers $x_k^j$ not need to be integer, then the set of all the possible strategies for player $j$
\begin{equation}
    \mathbb{S}_j= \bigg\{ \mathbf{x}_j \in \mathbb{R}^{+n} \bigg\vert \sum_{k=1}^n x_k^j=X_j \bigg\},
\end{equation}
where $\mathbb{R}^{+n}$ stands for the n-dimensional non-negative real set and $X_j$ is the total number of soldiers. Then each player choose a strategy and the payoff of the $j^{th}$ player can be calculated
\begin{equation}
  \$_j(X)= \sum_{k=1}^{n} sgn \left( x_k^j -x
 _{k,j}^{max} \right), 
\end{equation}
where $x
 _{k,j}^{max}$ is the highest number of soldiers among all the others players 
\begin{equation}
    x_{k,j}^{max}=\max\{x_k^1,x_{k}^2,...,x_{k}^{j-1},x_{k}^{j+1},..., x_{k}^N\},
\end{equation}
and 
\begin{equation}
  sgn(x)=\left\{
  \begin{array}{@{}ll@{}}
    +1, & \text{if}\ x\geq 0 \\
    0, & \text{if}\ x= 0 \\
    -1, & \text{if}\ x< 0
  \end{array}\right. .
\end{equation}

\section{The Quantum Game}\label{sec: quantumgame}
In order to define the quantum description of the game we choose the Hilbert space to be a direct product of $k$ soldier spaces, each corresponding to a player and the battlefield space
\begin{equation}
    \mathbb{H}= \mathbb{H}_S^1 \otimes \mathbb{H}_S^2 \otimes ... \otimes \mathbb{H}_S^N \otimes  \mathbb{H}_T,
\end{equation}
We use 2-dimensional soldier spaces $\mathbb{H}_S^j$ for any player $j$ and a n-dimensional battlefield space $\mathbb{H}_T$.
The game begins in an initial state $\ket{\psi_i}$ that belongs to the Hilbert space $\mathbb{H}$.
We represent our 2-dimensional space basis vectors as $\ket{0}_j$ and $\ket{1}_j$, where
\begin{equation}
\ket{0}_j = \begin{pmatrix} 1 \\ 0 \end{pmatrix}, \ \ \ \ \  \vert 1 \rangle_j = \begin{pmatrix} 0 \\ 1 \end{pmatrix} ,
\end{equation} 
and the battlefield's basis  $\{ \vert T_i \rangle \}$ with $i=1,2...n$ where,
\begin{equation}
\vert T_1 \rangle = \begin{pmatrix} 1 \\ 0 \\ \vdots \\ 0 \end{pmatrix}, \ \ \ \ \  \vert T_2 \rangle = \begin{pmatrix} 0 \\ 1 \\ \vdots \\0 \end{pmatrix},  \ \ \  \dots \ \ \  \vert T_n \rangle = \begin{pmatrix} 0 \\ 0 \\ \vdots \\ 1 \end{pmatrix}.
\end{equation}

The initial state of the system $\Psi_i$ is written as
\begin{equation}
\ket{\Psi_i} = \ket{0}_1 \otimes \ket{0}_2 \otimes ... \otimes \ket{0}_N \otimes \frac{1}{\sqrt{n}}\sum_{k=1}^n \ket{T_k},
\end{equation}
which we write in notation
\begin{equation}
\ket{\Psi_i} = \prod^{\otimes N}_{j=1} \ket{0}_j \otimes \frac{1}{\sqrt{n}}\sum_{k=1}^n \ket{T_k}.
\end{equation}
Blotto's games contain an usual condition which is $X_B\geq X_j$, i. e. Blotto has the highest total number of soldiers. The number of soldiers Blotto has i.e, $X_B$ is public information known to all players. To allocate troops, the $j^{th}$ player applies the unitary operator,
\begin{equation}
\hat{U}_j= \sum_{k=1}^n \mathbb{I}_1 \otimes ...\otimes \mathbb{I}_{j-1}\otimes \hat{R}(\lambda^j_k)\otimes \mathbb{I}_{j+1} \otimes...\otimes \mathbb{I}_N \otimes \hat{\Pi}_k ,
\end{equation}
where $j^{th}$ is the player, $\hat{\Pi}_k$ is the projector $\vert T_k \rangle \langle T_k \vert$, the operator $\hat{R}(\lambda^j_k)$ is the rotation matrix
\begin{equation}
\hat{R}(\lambda^j_k) = \begin{pmatrix} \cos(\lambda^j_k) & -\sin(\lambda^j_k) \\ \sin(\lambda^j_k) & \cos(\lambda^j_k)
\label{Crot}
\end{pmatrix} ,
\end{equation} 
and $\lambda^j_k$ is the angle which the vector $\ket{0}_j$ will be rotated. The rotation angle $\lambda^j_k$ is defined as
\begin{equation}\label{eq: lambda}
\lambda^j_k= \frac{\pi}{2} \frac{x^j_k}{X_B} ,
\end{equation} 
where $x^j_k$ stands for the number of $j^{th}$ player's troops in the $k^{th}$ battlefield. Therefore, each player rotates the $\ket{0}_j$ towards $\ket{1}_j$. This definition is important to depict the classical game, whoever has the highest projection on to $\ket{1}_j$ win a specific battlefield.The constraints are
\begin{equation}
    \sum_{k=1}^n x^j_k = X_j.
\end{equation}
This ensures that the rotation angle $\lambda^k_j \in [0,\frac{\pi}{2}]$. To introduce quantum strategies, we generalise the allocation operator for $j^{th}$ player to
\begin{equation}
\hat{U}_j= \sum_{k=1}^n \mathbb{I}_1 \otimes ...\otimes \mathbb{I}_{j-1}\otimes \hat{Q}(\lambda^j_k, \phi^j_k)\otimes \mathbb{I}_{j+1} \otimes...\otimes \mathbb{I}_N \otimes \hat{\Pi}_k ,
\end{equation}
where
\begin{equation}
\hat{Q}(\lambda^j_k , \phi^j_k) =  \begin{pmatrix} e^{i \phi^j_k} \cos(\lambda^j_k) & -\sin(\lambda^j_k) \\ \sin(\lambda^j_k) & e^{-i \phi^j_k}\cos(\lambda^j_k)
\label{Qrot}
\end{pmatrix} ,
\end{equation} 
where $\phi^j_k$ is a phase that represent a quantum resource and when $\phi^j_k=0$ one can easily verify $\hat{Q}(\lambda^j_k , 0)=\hat{R}(\lambda^j_k)$. To introduce entanglement in our initial state, we use an operator ${\hat J}$. In order to preserve the structure of a quantum game,
\begin{equation}\label{eq: psifinal}
    \ket{\Psi_f} = \hat{J}^\dagger \hat{U}_N ... \hat{U}_1 \hat{J}\ket{\Psi_i},
\end{equation}
the ${\hat J}$ must commute with all classical strategies,
\begin{equation}
\left[ \hat{J} ,  \sum_{k=1}^n \mathbb{I}_1 \otimes ...\otimes \mathbb{I}_{j-1}\otimes \hat{R}(\lambda^k_j)\otimes \mathbb{I}_{j+1} \otimes...\otimes \mathbb{I}_N \otimes \hat{\Pi}_k \right]=0.
\end{equation} 
A common procedure is to define the entanglement operator as
\begin{equation}
\hat{J}= e^{i \gamma \hat{A}/2},
\end{equation}
where $\gamma \in \left[0,\pi /2 \right]$ is a parameter that introduce the entanglement. Therefore, if the operator $\hat{A}$ satisfy $\hat{A}^{2n} = 1$, then
\begin{equation}
    \hat{J}=\cos(\gamma/2)\hat{\mathbb{I}}+i\sin(\gamma/2)\hat{A}.
\end{equation}
We choose $\hat{A}$ 
\begin{equation}
\hat{A}= (-1)^N
\begin{pmatrix} 0 & 1 \\ -1 & 0 \end{pmatrix}_{1} \otimes ... \otimes \begin{pmatrix} 0 & 1 \\ -1 & 0 \end{pmatrix}_{N} \otimes \begin{pmatrix} 
\pm i &  0 & \ldots & 0 \\  0 & \pm i & \ldots &  0\\ \vdots & \vdots & \ddots & \vdots \\ 0 & 0 & \ldots &\pm i\\
\end{pmatrix}_{n \times n} ,
\end{equation}
i.e, 
\begin{equation}
    \hat{A}= (-1)^N\prod^{\otimes N}_{j=1} \begin{pmatrix} 0 & 1 \\ -1 & 0 \end{pmatrix}_{j} \otimes \begin{pmatrix} 
\pm i &  0 & \ldots & 0 \\  0 & \pm i & \ldots &  0\\ \vdots & \vdots & \ddots & \vdots \\ 0 & 0 & \ldots &\pm i\\
\end{pmatrix}_{n \times n},
\end{equation}
which have an important condition: in the last matrix, at least one of the elements must have a different sign when compared to the others.
All the players play sequentially and the final state obtained at equation (\ref{eq: psifinal}), then we construct the density matrix
\begin{equation}
\rho = \ket{\Psi_f} \bra{\Psi_f}.
\end{equation}
To find the payoff of the $j^{th}$ player, partial trace is taken
\begin{equation}
    \rho_{j}= tr_X \rho,
\end{equation}
where X runs over all players except j, i.e, $X=\{1,2,..., j-1,j+1,..., N\}$. 
Therefore we perform a measurement
\begin{equation}
M_{k,j}= tr \left[\left( \ket{1}_j \bra{1}_j \otimes  \ket{T_k}  \bra{T_k}\right) \rho_{j}\right].
\end{equation}
For a given battlefield k, we define the quantity
\begin{equation}\label{eq: Mquantity}
M_{k,j}^{max}=\max\{M_{k,1},M_{k,2},...,M_{k,j-1},M_{k,j+1},...M_{k,N}\} ,
\end{equation}
for all n battlefields. The idea behind the definition of equation (\ref{eq: Mquantity}) is to obtain the highest value of the measurements $M_{k,i}$ for all $i$ players except $i=j$. This quantity is useful to define the payoff for the $j^{th}$ player
\begin{equation}\label{eq: payoff}
    \$_j = \sum_{k\neq j}^n sgn \left[M_{k,j}-M_{k,j}^{max}\right].
\end{equation}
The idea behind the definition of the payoff is to compare the measurement made by player $j$ with the highest value among all the others players.
\section{Results}\label{sec: results}
In this section we have shown examples in order to elucidate some features of the quantum game. In subsection \ref{subsec: classicalstrategies} we show a situation where all players have access only to classical strategies. In other hand, we allow enemy 3 to perform a quantum move in subsection \ref{subsec: quantummove}.
\subsection{Classical strategy}\label{subsec: classicalstrategies}
Consider an example with 3 players respectively fighting over 2 territories. Blotto has 6 soldiers, enemy 1 has 4 soldiers and enemy 2 has 3 soldiers.
Let Blotto choose the strategy (3,3), enemy 1 choose (3,1) and enemy 2 choose (0,3) as shown in Table (\ref{tab: classicalstrategies}).

\begin{table}[h!]
\begin{center}
\begin{tabular}{@{}llll@{}}
\toprule
\textbf{\begin{tabular}[c]{@{}l@{}}Battlefield $\rightarrow$\\ Players$\downarrow$\end{tabular}} & \textbf{Battlefield 1} & \multicolumn{2}{l}{\textbf{Battlefield 2}} \\ \midrule
\textbf{Blotto}                                                                                                                          & 3                      & \multicolumn{2}{l}{3}                      \\
\textbf{enemy 1}                                                                                                                         & 3                      & \multicolumn{2}{l}{1}                      \\
\textbf{enemy 2}                                                                                                                         & 0                      & \multicolumn{2}{l}{3}                      \\ \bottomrule
\end{tabular}
\caption{The allocation of soldiers for all players over 2 battlefields. Each number of soldier allocated to each battlefield has a correspondence with an angle, presented at equation (\ref{eq: lambda}).}
\label{tab: classicalstrategies}
\end{center}
\end{table}
In battlefield 1, both Blotto and enemy 1 tied whereas in Battlefield 2, both Blotto and enemy 2 tied.
Clearly, the payoffs for Blotto, enemy 1 and enemy 2 are $0$, $-1$ and $-1$ respectively.
We will now play the same strategies in quantum formalism. Blotto has maximum number of soldiers $X_B=6$. Thus the angles to be rotated in the corresponding qubits of the initial state are $(\frac{\pi}{4},\frac{\pi}{4})$, $(\frac{\pi}{4}, \frac{\pi}{12})$ and $(0,\frac{\pi}{4})$ by Blotto, enemy 1 and enemy 2 respectively. We tabulate the measurements in Table (\ref{tab:m-table}). Therefore, the payoffs can be calculated with the help of equation (\ref{eq: payoff}) which are $0$, $-1$ and $-1$, for Blotto, enemy 1 and enemy 2, respectively. Through this example, we verified that the classical payoffs agree with their counterpart classical strategies in the quantum game. In addition, one can calculate the measurements $M_{k,j}$ via 
\begin{equation}
    M_{k,j}= \frac{1}{n} \sin^2{\lambda^k_j},
\end{equation}
when all players are restricted to play with classical strategies
\begin{equation}
    \phi^k_j=0 \quad \forall \quad j\leq N \quad {\rm and} \quad k\leq n.
\end{equation}

\begin{table}[h!]
\begin{center}
\begin{tabular}{@{}llll@{}}
\toprule
\textbf{\begin{tabular}[c]{@{}l@{}}Battlefield $\rightarrow $ \\ Players $\downarrow$ \end{tabular}} & \textbf{Battlefield 1} & \multicolumn{2}{l}{\textbf{Battlefield 2}}              \\ \midrule
\textbf{Blotto}                                                                                                                          & $\frac{1}{4}$           & \multicolumn{2}{l}{$\frac{1}{4}$}                        \\
\textbf{enemy 1}                                                                                                                         & $\frac{1}{4}$           & \multicolumn{2}{l}{$\frac{1}{2} \sin^2{\frac{\pi}{12}}$} \\
\textbf{enemy 2}                                                                                                                         & 0                      & \multicolumn{2}{l}{$\frac{1}{4}$}                        \\ \bottomrule
\end{tabular}
\caption{The values of the measurements for the strategies presented at table \ref{tab: classicalstrategies}}
\label{tab:m-table}
\end{center}
\end{table}

\subsection{A quantum move}\label{subsec: quantummove}
In this subsection we shown some examples where the enemy 2 has access to quantum strategy. Through the following cases we consider the maximum entanglement which is $\gamma=\pi/2$. 
\subsubsection{First example}\label{subsubsec: firstex}
\begin{figure}
    \centering
    \includegraphics[width=\textwidth]{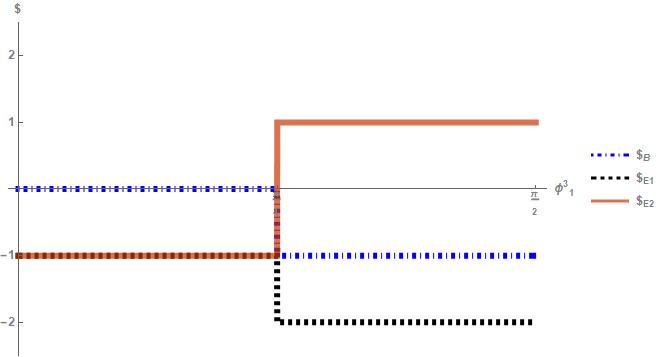}
    \caption{Plot of the payoff of all players. The blue dot-dashed line corresponds to the Blotto's payoff, the dotted black line to the enemy's 1 payoff and the red full line to the enemy's 2 payoff. In this case the quantum resource on the second battlefield is $\phi_2^3=0$. One can see that enemy 3 can have positive payoff for some values of $\phi_1^3$.}
    \label{fig: phi0}
\end{figure}

First we analyse the case which both Blotto and enemy 1 choose the same strategy presented at table (\ref{tab: classicalstrategies}). However enemy 2 will choose the strategy $x_1^3=0$, $x_2^3=3$, $\phi_2^3=0$ and $\phi_1^3$ for several values. The payoffs are plotted at figure (\ref{fig: phi0}) for each $\phi_1^3$ that belongs to the domain $\left[0,\pi/2\right]$, for all players. The enemy 2 can increase its own payoff if he choose a quantum strategy with $\phi_1^3>\pi/4$.

\subsubsection{Second example}
\begin{figure}
    \centering
    \includegraphics[width=\textwidth]{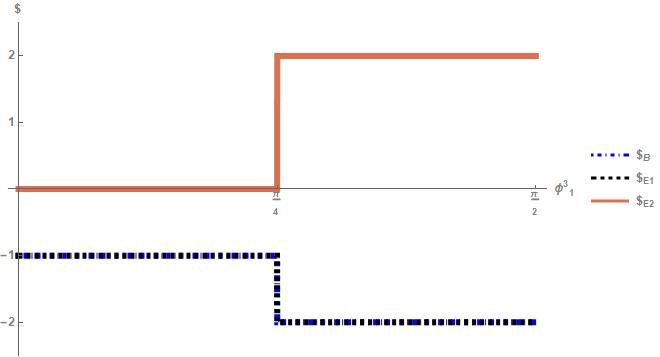}
    \caption{Plot of the payoff of all players. The blue dot-dashed line corresponds to the Blotto's payoff, the dotted black line to the enemy's 1 payoff and the red full line to the enemy's 2 payoff. }
    \label{fig: phivar}
\end{figure}

Another example, which is illustrated at figure (\ref{fig: phivar}), contains the same values used on the first example, sub-subsection (\ref{subsubsec: firstex}). However, we let $\phi_2^3$ vary. We observe no change of players' payoff for any values between $0<\phi_2^3<\pi/2$. In other words, the payoffs has a discontinuity jump at $\phi_2^3=0$, as one can see after the comparison between the figures (\ref{fig: phi0}) and (\ref{fig: phivar}). Therefore, considering both 
previous scenarios, it is better for the enemy 2 to choose values of quantum resources $0<\phi_2^3<\pi/2$ and $\phi_1^3>\pi/4$, which gives $\$_{3}=+2$ . 

\section{Conclusion}
We have presented a quantum version of the multiplayer Colonel Blotto game. Through our work we see that the payoff do not alter under the change of the order of the operators. This happens because the operators' strategies of different players commute with each other. We also find an great advantage for players who has access to quantum strategies.

\begin{acknowledgements}
Part of this work is supported by INSPIRE, Dept. of Science \& Technology, Govt. of India.
\end{acknowledgements}


\bibliographystyle{unsrt}
\bibliography{main}   

\end{document}